\documentclass{llncs}
\usepackage[utf8]{inputenc}
\usepackage[T1]{fontenc}
\usepackage{amssymb,amsmath,xspace,hyperref,graphicx,multirow}

 \newtheorem{cons}{Constraint}

\begin{document}

\pagestyle{empty}

\mainmatter

\title{Formalization of Complex Vectors in \\
Higher-Order Logic \thanks{We thank three referees for their insightful comments.
The final publication is available at http://link.springer.com.}}

\author{Sanaz Khan-Afshar  \and Vincent Aravantinos \and Osman Hasan \and Sofi\`{e}ne Tahar}
\institute{Dept. of Electrical \& Computer Engineering, Concordia University \\
1455 de Maisonneuve W., Montreal, Quebec, H3G 1M8, Canada\\
\email{\{s\_khanaf,vincent,o\_hasan,tahar\}@ece.concordia.ca}\\
\url{http://hvg.ece.concordia.ca}} \maketitle

\newcommand\indnt{\phantom{\ensuremath{\mathtt{\vdash\ }}}}
\newcommand\indntt{\indnt\indnt}
\newcommand\indnttt{\indntt\indnt}
\newcommand\C{\ensuremath{complex}\xspace}
\newcommand\R{\ensuremath{real}\xspace}
\newcommand\Cthree{\ensuremath{\mathtt{complex^3}}\xspace}
\newcommand\CN{\ensuremath{\mathtt{{complex^N}}}\xspace}
\newcommand\cfun{\ensuremath{\mathtt{cfun}}\xspace}
\newcommand\dst{\ensuremath{\mathtt{dest\_cart}}\xspace}
\newcommand\mk{\ensuremath{\mathtt{mk\_cart}}\xspace}
\newcommand\copofmat{cop\_of\_cmatrix\xspace}
\newcommand\cmatofop{cmatrix\_of\_cop\xspace}

\def\hol#1{$\mathtt{#1}$}
\def\vvec#1{\vec{#1}\,}
\def\rr{\vec{r}}
\def\vv{\vec{v}}
\def\ne{n_{\rm e}}
\def\ni{n_{\rm i}}
\def\no{n_{\rm 0}}
\def\me{m_{\rm e}}
\def\mi{m_{\rm i}}
\def\Te{T_{\rm e}}
\def\Ti{T_{\rm i}}
\def\Leftrightarrowdef{\Leftrightarrow}
\def\isincidentbasis{is\_incident\_basis\ }
\def\Xofbasis{x1\_of\_basis\ }
\def\Yofbasis{x2\_of\_basis\ }
\def\Zofbasis{x3\_of\_basis\ }

\def\emf{emf}
\let\odot=\cdot
\def\Lambda#1{\lambda #1\qdot}
\def\wofwave{\omega\_of\_w\ }
\def\eofwave{e\_of\_w\ }
\def\Exists#1{\exists #1\qdot}
\def\mapt{map\_triple\ }
\def\hofwave{h\_of\_w\ }
\def\planewave{plane\_wave\ }
\def\eqdef{=}
\def\NS{\negthinspace\negthinspace}
\def\isplanewaveatitf{is\_plane\_wave\_at\_interface\ }
\def\isplanewave{is\_plane\_wave\ }
\def\isvaliditf{is\_valid\_interface\ }
\def\isinplan{is\_in\_plane\ }
\def\kofwave{k\_of\_w\ }
\def\boundaryconditions{boundary\_conditions\ }
\def\Normalofplan{norm\_of\_plane}
\def\normalofplan{\Normalofplan\ }
\def\forceindent{\ \ }
\def\vdashdef{\vdash}
\def\qdot{.\ }
\def\Forall#1{\forall #1\qdot}
\def\isposonbasis{is\_positive\_orthonormal\_basis\ }
\def\planofitf{plane\_of\_interface\ }
\def\aresymetric{are\_sym\_wrt\ }
\def\vectorsine{vector\_sine\ }
\def\vecnorm#1{\parallel\negthinspace #1\negthinspace\parallel}
\def\temode{te\_mode\ }
\def\tmmode{tm\_mode\ }
\def\boundaryconds{boundary\_conditions\ }
\def\eofemf{e\_of\_emf\ }
\def\hofemf{h\_of\_emf\ }

  \begin{abstract}
Complex vector analysis is widely used to analyze continuous systems in many disciplines, including physics and engineering. In this paper, we present a
higher-order-logic formalization of the complex vector space to facilitate conducting this analysis within the sound core of a 
theorem prover: HOL Light. Our definition of complex vector builds upon the
definitions of complex numbers and real vectors. This extension allows us to extensively benefit from the already verified theorems based on complex analysis
and real vector analysis. To show the practical usefulness of our library 
we adopt it to formalize electromagnetic fields and to prove the law of reflection for the planar waves.

  \end{abstract}

  \section{Introduction} \label{sec:intro}
 
 Vector analysis is one of the most useful branches of mathematics; a highly scientific field that is used in practical problems arising in engineering and applied sciences. 
Not only the real vectors but the \emph{complex} vectors are a convenient tool to describe natural and physical phenomena, including waves. They are thus used in fields as diverse as optics, electromagnetics, quantum mechanics, nuclear physics, aerospace, and communications. 
Therefore, a concrete development of (real and complex) vector analysis by formal methods can be a huge asset in the development and application of formal methods in connection with a variety of disciplines, e.g., physics, medicine, control and signal processing.

 The theory of vector analysis is formalized in many theorem provers, e.g., HOL Light \cite{harrison_13hol}, Isabelle/HOL \cite{isabelleVEC}, PVS \cite{PVS}, and Coq \cite{CoqLinearAlgebra}. 
However, these works are either limited to real vector analysis or
    provide very abstract formalizations which are useful for the formalization
    of mathematics but lack many notions required for applied sciences. 
    For instance,  Coq \cite{CoqLinearAlgebra} provides a general, axiomatic,
  formalization of linear algebras. However, this work lacks many notions
  that are useful for applications.
In PVS \cite{PVS}, real vector spaces are formalized but the work does not support complex vectors. 
Another example is the Isabelle/HOL \cite{isabelleVEC} with a library of real vector analysis ported from the formalization of multivariate analysis available in the HOL Light theorem prover \cite{harrison_13hol}. To the best of our knowledge, the only work addressing complex vectors is 
\cite{NFMYousri}, where the authors present a formalization of complex vectors, using the concept of complex function spaces in HOL Light. However, their formalization is very abstract and is focused on infinite dimension linear spaces: those definitions and properties which are only meaningful in a finite dimension are not addressed in the formalization presented in \cite{NFMYousri}, e.g., cross product.

In this paper we present an alternate formalization of complex vectors using the HOL Light theorem prover. 
HOL Light is an interactive theorem prover which is based on classical higher-order logic with axioms of infinity, extensionality, and choice in the form of Hilbert's \hol{\epsilon} operator \cite{harrison_09hol}. HOL Light uses functional programming language Objective CAML (OCaml) as both the implementation and interaction language. 
This theorem prover has been particularly successful in verifying many challenging mathematical theorems by providing formal reasoning support for different mathematical theories, including real analysis, complex analysis and vector calculus. 
The main motivation behind choosing HOL Light for the formalization of complex vector analysis is the availability of rich libraries of multivariate analysis, e.g., complex analysis \cite{harrison_07} and Euclidean space \cite{harrison_13hol}. 

Our formalization of complex vectors is inspired by the concept of bivectors, originally introduced by the famous American physicist J. William Gibbs, in his famous pamphlet ``Elements of Vector Analysis'' \cite{gibbs_1884}. We adopt a great part of definition and properties of vector analysis and extend them to our formalization of complex vectors. Then, we prove many of complex vectors properties by introducing componentwise operators, and inheriting properties of complex analysis from HOL Light multivariate libraries \cite{harrison_07}. Our formalization thus overcomes the limitations of \cite{NFMYousri} by providing the support of finite vector spaces, in two of the most widely used representations of vectors in applied sciences: vector of complex numbers and bivectors. 
In general, the subject of vector analysis can be sub-divided into three distinct parts \cite{gibbs_1884}, i.e, the algebra of vectors, the theory of the linear vector function, and the differential and integral calculus of vector functions. In this paper, we mainly focus on the first two parts.

The rest of the paper is organized as follows: In Section \ref{sec:operators} we introduce two sets of operators, which are extensively used in our formalization to make use of the multivariate libraries of HOL Light. Section \ref{sec:CVS} presents
  our formalization of complex vector algebra followed by the formalization of linearity and infinite summation of complex vector functions. Finally, Section \ref{sec:application} provides an application that illustrates the usage of our current development of complex vectors by 
  the formalization of some basics of electromagnetics: In particular, the law of reflection and the law of plane of incidence for plane electromagnetic waves have been verified.
   
All the source codes corresponding to the work presented in this paper 
are available at:  {\url{http://hvg.ece.concordia.ca/projects/optics/cxvec/}.

\section{Complex Vectors vs. Bivectors}
\label{sec:operators}
Complex vectors, by definition, are vectors whose components are complex numbers. 
One approach to formalize complex vectors is as a pair of two real vectors, very similar to the definition of bivectors by J. William Gibbs \cite{gibbs_1884}. 
Adopting this approach, first, we instantly inherit all the topological and analytic apparatus of Euclidean Space, described in \cite{harrison_13hol}, for real and imaginary parts of complex vectors. Next, we can adopt the approach used for developing complex analysis based on real analysis \cite{harrison_07} to extend our libraries for complex vector analysis. 

However, in many analytical problems, we need to have access to each element of complex vectors as a complex number. For instance, in the standard definition of complex vector derivative, derivative exists if and only if a function of complex variable is  complex analytic, i.e., if it satisfies the Cauchy-Riemann equations \cite{LePage_80}.  However, this condition is very strong and in practice, many systems, which are defined as a non-constant real-valued functions of complex variables, are not complex analytic and therefore are not differentiable in the standard complex variables sense. In these cases, it suffices for the function to be only differentiable in directions of the co-ordinate system, i.e., in case of cartesian co-ordinate systems, for the $x$, $y$, and $z$. In these cases, it is preferred to define the complex vector as a vector of complex numbers.

In order to have the advantages of both approaches, we define a set of operators which makes a one to one mapping between these two representations of complex vectors.

First, we formalize the concept of vector operators, with no restriction on the data type\footnote{ In order to improve readability, occasionally, HOL Light statements are written by mixing HOL Light script notations and pure mathematical notations.}:

\begin{definition}[Unary and Binary Componentwise Operators]\ \label{def:operators} \vspace{.15cm}\\
\hol{\vdash\ \Forall {(k:A)} vector\_const\ k = \ lambda\ i.\ k\ : A^N} 	\vspace{.08cm}	\\
\hol{\vdash\ \Forall {(f:A\to B)(v:A^N)} vector\_map\ f\ v = \ lambda\ i.\ f(v\$ i) : B^N} 	\vspace{.08cm}	\\
\hol{\vdash\ \Forall {(f:A\to B \to C)(v1:A^N)(v2:B^N)} }\\
\hol{\hspace{3cm}vector\_map2\ f\ v1\ v2 = \ lambda\ i.\ f(v1\$ i) (v2\$ i): C^N} 		
		\end{definition}
\noindent where  $\mathtt{v\$i}$ returns the $\mathtt i^{th}$ component of $\mathtt v$ and $\mathtt{lambda\ i.\ f(v\$ i)}$ applies the unary operator \hol f on each component of $\mathtt v$ and returns the vector which has \hol{f(v\$n)} as its nth component, i.e., $\mathtt{f (v\$ 1)}$ as its first component,
$\mathtt{f(v\$ 2)}$ as its second component, and so on. 

In Section \ref{sec:CVS}, we will show how to verify all the properties of ``componentwise'' operations of one (complex) vector space by its \emph{counterpart field} (complex numbers) using Definition \ref{def:operators}. For instance, after proving that: \vspace{.2cm}\\
\noindent  \hol{\vdash\ \Forall{i\ f\ v} (vector\_map\ f\ v )\$i\ =\ f (v\$i)} \vspace{.25cm}\\
 \noindent it is trivial to prove that real vector negation is \hol {vector\_map} of real negation: \vspace{.2cm}\\
\noindent \hol{\vdash\ (--): real^N \to real^N = vector\_map ((--): real \to real)}. \vspace{.25cm}

Next, we extract the real and imaginary parts of a complex vector with type \hol{complex^N} as two \hol{real^N} vectors, by \hol{cvector\_re} and \hol{cvector\_im}, respectively, and also import the real and imaginary parts of a complex vector into its original format by \hol{complex\_vector}, as follows:

\begin{definition}[Mapping between \hol{complex^N} and \hol{real^N \times real^N}] \label{def:mapping} \vspace{.15cm}\\ 
\hol {\vdash cvector\_re  = vector\_map\ Re}  \vspace{.08cm}\\
\hol {\vdash cvector\_im = vector\_map\ Im}  \vspace{.08cm} \\
\hol {\vdash  \Forall{v1\ v2} complex\_vector\ (v1,v2) = }\\
\hol{\hspace{2cm}vector\_map2\ (\lambda\ x\ y. Cx\ x + ii * Cx\ y)\ v1\ v2}
\end{definition}

Finally, we formally define a bijection between complex vectors and real vectors with even size, as follows: 
\begin{definition}[ Flatten and Unflatten] \\
\hol{\vdash \Forall v flatten \ (v:complex^N) : real^{(N,N)finite\_sum} =}\\
\hol{\hspace{5cm}  pastecart\ (cvector\_re\ v)\ (cvector\_im\ v)}\vspace{.1cm}\\
\hol{\vdash \Forall v  unflatten\ (v:real^{(N,N)finite\_sum}) : complex^N  =}\\
\hol{\hspace{5cm}  complex\_vector\ (fstcart\ v, sndcart\ v)}
\end{definition}  
\noindent where type \hol{:real^{(N,N)finite\_sum}} refers to a real vector with size \hol{N+N}. All three functions \hol{pastecart}, \hol{fstcart}, and \hol{sndcart} are HOL Light functions. The function \hol{flatten}, takes a complex vector with size \hol N and returns a real vector with size \hol {N+N} in which the first \hol N elements provide the real part of the original vector and the second \hol N elements provide the imaginary part of the original complex vector. The \hol{unflatten} is an inverse function of \hol{flatten}. The most important properties of \hol{flatten} and \hol{unflatten} are presented in Table \ref{table:flatten}. The first two properties in Table \ref{table:flatten}, i.e., Inverse of Flatten and Inverse of Unflatten, guarantee that \hol{flatten} and \hol{unflatten} are bijective. 
\begin{table}[h]
			\renewcommand{\arraystretch}{1.2}
			\begin{center}{
			\begin{tabular}{ll}
			\textbf{Property} & \textbf{Formalization}  \ \\
			\hline
								Inv. of Flatten
									& \hol{\vdash unflatten\ o\ flatten = I :complex^N \to  complex^N}
									 \ \\
								Inv. of Unflatten
									& \hol{\vdash flatten\ o\ unflatten = I :real^{(N,N)finite\_sum} \to real^{(N,N)finite\_sum}}
									\ \\
									Flatten map
									& \hol{\vdash\ \Forall {f\ g} f = vector\_map\ g\ \Rightarrow\ } \\
								\ & \hol {\hspace{.4cm} \Forall x  flatten (vector\_map\ f\ x) = vector\_map\ g\ (flatten\ x)}
									 \ \\
									Flatten map2
									& \hol{\vdash\ \Forall {f\ g} f = vector\_map2\ g\ \Rightarrow\ } \\
								\ & \hol {\hspace{.4cm} \Forall {x\ y}  flatten (vector\_map2\ f\ x\ y) = }\\
								\ & \hol {\hspace{2cm} vector\_map2\ g\ (flatten\ x) (flatten\ y)}
				\end{tabular}}
			\end{center}
			\caption{Mapping Complex Vectors and Real Vectors}
						\label{table:flatten}
		\end{table}
The second two properties ensure that as long as an operator is a map from \hol{complex^N} to \hol{real^N}, applying this operator in the \hol{complex^N} domain, and flattening the result will give the same result as flatten operand(s) and applying the operation in the \hol{real^N} domain. This property is very helpful to prove componentwise properties of complex vectors from their counterparts in real vectors analysis.

  \section{Complex Vector Algebra}
  \label{sec:CVS}

The first step towards the formalization of complex vector algebra is to formalize \emph{complex vector space}. 
Note that the HOL Light built-in type of vectors does not represent, in general, elements of a vector space. Vectors in HOL Light are basically lists whose length is explicitly encoded in the type.
Whereas a vector space is a set $S$ together with a field $F$ and two binary operators that satisfy eight axioms of vector spaces\cite{Tallack_70} (Table \ref{table:VSA}).
Therefore, we have to define these operators for $\mathtt{{complex^N}}$ and prove that they satisfy the linear space axioms. 

\subsection*{Vector Space}
Now, it is easy to formally define ``componentwise'' operations of complex vectors using their counterparts in complex field, including addition and scalar multiplication: 

		\begin{definition}[Arithmetics over $\mathtt{{complex^N}}$]\ \vspace{.2cm}
			\label{def:cvector_add}
			\label{def:cvector_smul}
			\label{def:cvector_zero} \\
		\hol{\vdash\ cvector\_add = vector\_map2\ (+: complex \to complex ) } \\
		\hol{\vdash\ cvector\_mul\ (a:{complex}) = vector\_map\ (( *:complex \to complex)\ a) } 
	\end{definition}					 
		\begin{table}[h]
			\renewcommand{\arraystretch}{1.2}
			\begin{center}{
			\begin{tabular}{ll}
			\textbf{Property} & \textbf{Formalization}  \ \\
			\hline
								 Addition associativity
									& \hol{\vdash\ \forall\ u\ v\ w.\ u + v + w = (u + v) + w  } 
									 \ \\
									Addition commutativity
									& \hol{\vdash\ \forall\ u\ v.\ u + v = v + u}  
									\ \\
									Addition unit
									& \hol{\vdash\ \forall\ u.\ u + cvector\_zero = u}
									 \ \\
									 Addition inverse
									& \hol{\vdash\ \forall\ u.\ u + (--u) = cvector\_zero}
									 \ \\
									Vector distributivity
									& \hol{\vdash\ \forall\ a\ u\ v.\ a\ \%\ (u + v) = a\ \%\ u + a\ \%\ v    }
									 \ \\
									Scalar distributivity
									& \hol{\vdash\ \forall\ a\ b\ u.\ (a + b)\ \%\ u = a\ \%\ u + b\ \%\ u }            
									 \ \\
									Mul. associativity
									&\hol{\vdash\ \forall\ a\ b\ u.\ a\ \%\ b\ \%\ u = (a * b)\ \%\ u    }            
									 \ \\
								 Scalar mul. unit
								 & \hol{\vdash\ \forall\ u.\ Cx(\&1)\ \% \ u = u  }
				\end{tabular}}
			\end{center}
			\caption{Vector Space Axioms for Complex Vectors}
						\label{table:VSA}
		\end{table}
    We developed a tactic, called \texttt{CVECTOR\_ARITH\_TAC}, which is mainly adapted from \texttt{VECTOR\_ARITH\_TAC} \cite{harrison_13hol} and is able to prove simple arithmetics properties of complex vectors automatically.
Using this tactic we prove all eight axioms of vector spaces, indicated in Table \ref{table:VSA}, plus many other basic but useful facts. 
In Table \ref{table:VSA}, the symbol (\hol \&) and the function (\hol{Cx}) are HOL Light functions for typecasting from \hol{integer} to \hol{real} and from \hol{real} to \hol{complex}, respectively. The symbol (\hol \%) and (\hol {--}) are overloaded by scalar multiplication and complex vector negation, respectively, and \hol {cvector\_zero} is a complex null vector. The negation and \hol {cvector\_zero} are formalized using componentwise operators \hol{vector\_map} and \hol{vector\_const} (Definition \ref{def:operators}), respectively.

\subsection*{Vector Products}

Two very essential notions in vector analysis are \emph{cross product} and \emph{inner product}.

The cross product between two vectors $u$ and $v$ of size $3$ is classically defined as $(u_2v_3-u_3v_2,u_3v_1-u_1v_3,u_1v_2-u_2v_1)$.
This is formalized as follows\footnote{The symbol ``$\times$'' indicates \hol{ccross} in our codes.}: 
\begin{definition}[Complex Cross Product]   \label{def:ccross} \vspace{.2cm}\\
      $\mathtt{\vdash\ \forall u\ v:{\mathtt{complex^3}}.\  u \times v :{\mathtt{complex^3}} =}$\ \\
      $\mathtt{\indnt vector\ [u\$2 *v\$3 - u\$3 * v\$2;\ u\$3 * v\$1 - u\$1 * v\$3;\ u\$1 * v\$2 - u\$2 * v\$1] }$
    \end{definition} \noindent where \texttt{vector} is a HOL Light function taking a list as input and returning a vector.
Table \ref{table:cross} presents some of the properties we proved about the cross product.
\begin{table}
			\renewcommand{\arraystretch}{1.2}
			\begin{center}{
			\begin{tabular}{ll}
				\textbf{Property} & \textbf{Formalization}\ \\\hline
				Left zero & \hol{\vdash \forall\ u.\ cvector\_zero \times u = cvector\_zero}\ \\
				Right zero &\hol{\vdash \forall\ u.\ u \times cvector\_zero = cvector\_zero}\ \\
				Irreflexivity & \hol{\vdash \forall\ u.\ u \times u = cvector\_zero}\ \\
				Asymmetry & \hol{\vdash \forall\ u\ v. -- (u \times v) = v \times u}\ \\
				Left-distributivity over addition & \hol{\vdash\ \forall\ u\ v\ w.\ (u + z) \times w = u \times w + z \times w}\ \\
				Right-distributivity over addition &\hol{\vdash \forall\ u\ v\ w.\ u \times (v + w) = u \times v + u \times w}\ \\
			 	Left-distributivity over scalar mul. & \hol{\vdash \forall\ a\ u\ v.\ (a \% u) \times v = a \% (u \times v) }\ \\
			 	Right-distributivity over scalar mul. &\hol{\vdash \forall\ a\ u\ v.\ u \times (a \% v) = a \% (u \times v)}\ 
				\end{tabular}
			}
			\end{center}
			\caption{Cross Product Properties}
						\label{table:cross}
		\end{table}
    
    The inner product is defined for two complex vectors $u$ and $v$ of dimension $N$ as $\sum_{i=1}^N u_i \overline v_i$,
    where $\overline{x}$ denotes the complex conjugate of $x$.
    This is defined in HOL Light as follows:
    \begin{definition}[Complex Vector Inner Product]\ \vspace{.2cm}\\
      \label{def:cdot}
      \hol{\vdash\ \Forall {(u:{complex^N})\ (v:{complex^N})}}\\
      \hol{\hspace{3cm}u\ cdot\ v = vsum\ (1..dimindex(:N))\ (\lambda i.\ u\$i * cnj(v\$i))}
    \end{definition}
   \noindent where \texttt{cnj} denotes the complex conjugate in HOL Light, and \texttt{vsum s f} denotes $\sum_{\mathtt{x\in s}} \mathtt {f\ x}$,
    \texttt{dimindex s} is the number of elements of \texttt s,
    and $\mathtt{(:N)}$ is the set of all inhabitants of the type \texttt N 
    (the ``universe'' of \texttt N, also written \texttt{UNIV}).

    Proving properties for the \emph{inner product space}, presented in Table \ref{table:IPS}, are quite straightforward except for the positive definiteness, which involves inequalities. The inner product of two complex vector is a complex number. Hence, to prove the positive definiteness, we first prove that \hol {\vdash \Forall x real(x\ cdot\ x)}, where \hol{real} is a  HOL Light predicate which returns true if the imaginary part of a complex number is zero. Then, by introducing function \hol{real\_of\_complex}, which returns a complex number with no imaginary part as a real number, we formally prove the positive definiteness. This concludes our formalization of complex inner product spaces. 
  
 		\begin{table}
			\renewcommand{\arraystretch}{1.2}
			\begin{center}{
			\begin{tabular}{ll}
			\textbf{Property} & \textbf{Formalization}  \ \\
			\hline
								Conjugate Symmetry
									& \hol{\vdash \Forall {x\ y} x\ cdot\ y = cnj (y\ cdot\ x)}
									 \ \\
									Linearity (scalar multiplication)
									& \hol{\vdash \Forall {c\ x\ y}(c\ \%\  x)\ cdot\ y = c * (x\ cdot\ y)}
								        \\ 
									Linearity (vector addition)									
									&  \hol{\vdash \Forall {x\ y\ z}(x + y)\ cdot\ z = (x\ cdot\ z) + (y\ cdot\ z)}
									\ \\
									Zero length 
									& \hol{\vdash \Forall x x\ cdot\ x = Cx(\&0)  \Leftrightarrow\ x = cvector\_zero}
									\ \\
									 Positive definiteness								
									& \hol{\vdash \Forall x \&0\leq  real\_of\_complex (x\ cdot\ x) }
				\end{tabular}}
			\end{center}
			\caption{Inner Product Space}
						\label{table:IPS}
		\end{table}

Norm, orthogonality, and the angle between two vectors are mathematically defined using the inner product.
Norm is defined as follows: 

\begin{definition}[Norm of Complex Vectors] \\
\hol {\vdash cnorm  = sqrt\ o\ cnorm2} 
\end{definition}
\noindent where \hol{\vdash \Forall x cnorm2\ x =  real\_of\_complex (x\ cdot\ x)}. Then the \hol{norm} is overloaded with our new definition of \hol{cnorm}.

We also define the concept of orthogonality and collinearity, as follows: 
\begin{definition}[Orthogonality and Collinearitiy of Complex Vectors] \vspace{.2cm}\\
\hol {\vdash \Forall{x\ y} corthogonal\ x \ y \Leftrightarrow x\ cdot\ y = Cx(\&0)} \\
		\hol { \vdash \Forall {x\ y}collinear\_cvectors\ x\ y\ \Leftrightarrow\ \Exists a (y = a \% x)\ \vee\ (x = a \% y)}
\end{definition}

Next, the angle between two complex vectors is formalized just like its counterpart in real vectors. Obviously, defining the vector angle between any complex vector and \hol{cvector\_zero} as \hol{Cx(pi / \& 2)} is a choice, which is widely used and accepted in literature. Note that in a very similar way we define Hermitian angle and the real angle \cite{cangle_01}. 
\begin{definition}[Complex Vector Angle] \vspace{.2cm}\\
\hol {\vdash \Forall{x\ y:complex^N}}\\
\hol{ cvector\_angle\ x\ y =  if\ x =  cvector\_zero\ \vee\ y = cvector\_zero\ } \\
\hol{\hspace{2cm}then\ Cx(pi / \& 2)}\\
\hol{\hspace{2cm}else\ cacs ( (x\ cdot\ y) / Cx(norm\ x\ *\ norm\ y) )}
\end{definition}

In Table \ref{table:products}, some of the properties related to norm and analytic geometry are highlighted.

		\begin{table}
			\renewcommand{\arraystretch}{1.2}
			\begin{center}{
			\begin{tabular}{ll}
			\textbf{Property} & \textbf{Formalization}  \ \\
			\hline
						Cross product \& collinearity			
						&  \hol{\vdash \Forall {x\ y} x \times y = cvector\_zero\  \Leftrightarrow\ }\\
						& \hspace{3.8cm}\hol{collinear\_cvectors\ x\ y}
									\ \\
						Cauchy-Schwarz inequality
						& \hol {\vdash \Forall {x\ y} norm (x\ cdot\ y) \leq norm\ x * norm\ y}
									 \ \\
						Cauchy-Schwarz equality
						& \hol {\vdash \Forall {x\ y} collinear\_cvectors\ x\ y \  \Leftrightarrow\  } \\
						\ &\hspace{2.5cm} \hol{norm (x\ cdot\ y) = norm\ x * norm\ y}
									 \ \\
						Triangle inequality
						& \hol {\vdash \Forall {x\ y} norm(x + y) \leq norm\ x + norm\ y}
								        \\							 				
						Pythagorean theorem
						&  \hol{\vdash \Forall {x\ y}  corthogonal\ x\ y \Leftrightarrow\ }\\
						& \hspace{2cm}\hol{cnorm2 (x+y) = cnorm2\ x + cnorm2\ y}
									\ \\
						Dot product and angle
						&  \hol{\vdash \Forall {x\ y} x\ cdot\ y = Cx (norm\ x * norm\ y)* }\\ 
						\ & 
						\hol{\hspace{3.7cm}ccos(cvector\_angle\ x\ y)}						
						\ \\
						Vector angle range
						&  \hol{\vdash \Forall {x\ y}	 \neg collinear\_cvectors\ x\ y \Rightarrow }\\
						\ & 
						\hspace{2.7cm}\hol{\&0< Re (cvector\_angle\ x\ y)\ \wedge\ }\\
						\ & 
						\hspace{3.1cm}\hol{Re(cvector\_angle\ x\ y)< pi}			
				\end{tabular}}
			\end{center}
			\caption{Highlights of properties related to vector products}
						\label{table:products}
		\end{table}

			We also define many basic notions of linear algebra, e.g., the canonical basis of the vector space,
		i.e., the set of vectors $(1,0,0,0,\dots)$, $(0,1,0,0,\dots)$, $(0,0,1,0,\dots)$, and so on.
		This is done as follows:
    \begin{definition}
      \label{def:cbasis}
      \hol{\vdash\ \forall k.\ cbasis\ k = vector\_to\_cvector\ (basis\ k) : complex^N}
    \end{definition}
		With this definition $\mathtt{cbasis\ 1}$ represents $(1,0,0,0,\dots)$,
		$\mathtt{cbasis\ 2}$ represents $(0,1,0,0,\dots)$, and so on.
		
		Another essential notion of vector spaces is the one of \emph{matrix}.
    Matrices are essentially defined as vectors of vectors:
    a $M\times N$ matrix is formalized by a value of type $\mathtt{(complex^N)^M}$.
		Several arithmetic notions can then be formalized over matrices, again using the notions of operators, presented in Definition \ref{def:operators}. Table \ref{table:matrix} presents the formalization of complex matrix arithmetics. The formal definition of arithmetics in matrix is almost identical to the one of complex vectors, except for the type. Hence, in Table \ref{table:matrix}, we specify the type of functions, where \hol{\mathbb{C}} represents the type \hol{complex}.

		\begin{table}
			\renewcommand{\arraystretch}{1.2}
			\begin{center}{
			\begin{tabular}{lll}
			\textbf{Operation} &  \textbf{Types} & \textbf{Formalization}  \ \\
			\hline
								Negation
									&  \scriptsize{\hol{: {\mathbb{C}^N}^M \to {\mathbb{C}^N}^M} }
									& \hol{\vdash cmatrix\_neg = vector\_map\ (--)}
									\ \\
								Conjugate
									&\scriptsize{\hol {: {\mathbb{C}^N}^M \to {\mathbb{C}^N}^M} }
									& \hol{\vdash cmatrix\_cnj = vector\_map\ (cvector\_cnj)}
									\ \\
								Addition
									&\scriptsize{\hol {: {\mathbb{C}^N}^M \to {\mathbb{C}^N}^M \to {\mathbb{C}^N}^M} }
									& \hol{\vdash cmatrix\_add = vector\_map2\ (+)}
									\ \\
								Scalar Mul.
									&\scriptsize{\hol {:  \mathbb{C} \to {\mathbb{C}^N}^M \to {\mathbb{C}^N}^M} }
									& \hol{\vdash cmatrix\_smul = vector\_map\ o\ (\%)}
									\ \\
								Mul.
									&\scriptsize{\hol {: {\mathbb{C}^N}^M \to {\mathbb{C}^P}^N \to {\mathbb{C}^P}^M}} 
									& \hol{\vdash \Forall {m_1\ m_2}  cmatrix\_mul = }
									\\
								\ & \ 
								       & \hol{\hspace{.3cm}lambda\ i\ j. vsum(1.. dimindex(:N))\ (\lambda k. m_1\$i\$k * m_2\$k\$j)}
				\end{tabular}}
			\end{center}
			\caption{Complex Matrix Arithmetic}
						\label{table:matrix}
      \vspace*{-0.2cm}
		\end{table}

Finally, we formalize the concepts of summability and infinite summation, as follows: 
\begin{definition}[Summability and Infinite Summation] \vspace{.2cm}\\
\hol {\vdash \Forall {(s:num\to bool) (f:num\to complex^N)}} \\
 \hol {  \hspace{.3cm} csummable\ s \ f \Leftrightarrow   summable\ s (cvector\_re\ o\ f)\ \wedge\ }\\
 \hol { \hspace{3.2cm}\ summable\ s\ (cvector\_im\ o\ f) } \vspace{.1cm}\\
\hol {\vdash \Forall {(s:num\to bool) (f:num\to complex^N)}} \\
 \hol { \hspace{.3cm} cinfsum\ s\ f =  vector\_to\_cvector\ (infsum\ s\ (\lambda x. cvector\_re\ (f\ x)))  } \\
\hol { \hspace{2.3cm}\ +\  ii \% vector\_to\_cvector (infsum\ s\ (\lambda x.cvector\_im\ (f\ x)))} 
\end{definition}

We, again, prove that summability and infinite summation can be addressed by their counterparts in real vector analysis. 
Table \ref{table:sum} summarizes the key properties of summability and infinite summation, where the predicate \hol{clinear\ f} is \hol{true} if and only if the function \hol{f: complex^M \to complex^N} is linear. As it can be observed in Table \ref{table:sum}, to extend the properties of summability and infinite summation of real vectors to their complex counterparts, the two functions of \hol{flatten} and \hol{unflatten} are used. When we \hol{flatten} a complex vector, the result would be a real vector which can be accepted by \hol{infsum}. Since summation is a componentwise operand and, as presented in Table \ref{table:flatten},  \hol{unflatten} is the inverse of  \hol{flatten}, \hol{unfaltten}ing the result of \hol{infsum} returns the desired complex vector.

		\begin{table}[h]
			\renewcommand{\arraystretch}{1.2}
			\begin{center}{
			\begin{tabular}{ll}
			\textbf{Properties} & \textbf{Formalization}  \ \\
			\hline
								Linearity
									& \hol {\vdash \Forall f clinear\ f \Leftrightarrow\  linear\ (flatten\ o\ f\ o\ unflatten)} 
									\ \\
						Summability 
						& \hol {\vdash \Forall {s\ f }csummable\ s \ f \Leftrightarrow\ summable\ s\ (flatten\ o\ f)} \\
					Infinite Summation
						&  \hol{\vdash \Forall {s\ f} csummable\ s\ f \Rightarrow} \\
						\ & \hol { \hspace{.5cm} (cinfsum\ s\ f  = unflatten\ (infsum\ s\ (flatten\ o\ f)))} \\
				\end{tabular}}
			\end{center}
			\caption{Linearity and summability of complex vector valued functions.}
						\label{table:sum}
      \vspace*{-0.2cm}
		\end{table}
\bigskip
In summary, we successfully formalized 30 new definitions and proved more than 500 properties in our libraries of complex vectors. 
The outcome of our formalization is a set of libraries that is very easy to understand and to be adopted for the formalization of different applications, in engineering, physics, etc. 	
Our proof script consists of more than 3000 lines of code and took about 500 man-hours for the user-guided reasoning process.

\subsection*{Infinite Dimension Complex Vector Spaces}
    \label{sec:cfun}
As mentioned earlier in Section \ref{sec:intro}, in \cite{NFMYousri}, the infinite-dimension complex vector spaces are formalized using function spaces. 
However, this definition brings unnecessary complexity to the problems in the finite space. As a result, we chose to develop our library of complex vectors then prove that there exist an isomorphism between our vector space and a finite vector space developed based on the work presented in \cite{NFMYousri}.

The complex-valued functions, in \cite{NFMYousri}, are defined as functions from an arbitrary set to \hol{complex}, i.e.,
    values of type \hol{cfun=A\to complex}, where \hol A is a type variable.
In general, complex vector spaces can be seen as a particular case of complex function spaces. 
		For instance, if the type \hol A in \hol{A\to complex} is instantiated with a
		finite type, say a type with three inhabitants \hol{one,two,three},
		then functions mapping each of \hol{one,two,three} to a complex number
		can be seen as $3$-dimension complex vectors. 		
So, for a type \hol t with \hol n inhabitants,
		there is a bijection between complex-valued functions of domain \hol t
		and complex vectors of dimension \hol n.
		
		This bijection is introduced by Harrison \cite{harrison_13hol}, defining the type
		constructor \hol{(A)finite\_image} which has a finite
		set of inhabitants. If \hol A is finite and has
		\hol n elements then \hol{(A)finite\_image} also has \hol n elements.
		Harrison also defines two inverse operations, as follows:\vspace{.2cm}\\
		\hol{\vdash mk\_cart :((A)finite\_image\to B)\to B^A} \vspace{.1cm}\\
		\hol{\vdash dest\_cart :B^A\to((A)finite\_image\to B)}\vspace{.2cm}\\
    where \hol{B^A} is the type of vectors with as many coordinates as the number of elements in \hol A. 
		
		By using the above inverse functions, we can transfer complex vector functions, for instance, ${\mathtt{cfun}}$ addition
		to the type \hol{complex^N} as follows: \vspace{.2cm}\\
    \hol{v+_{complex^N} w=mk\_cart\ (dest\_cart\ v+_{\mathtt{cfun}}\ dest\_cart\ w)}\vspace{.2cm}\\
    \noindent where the indices to the \hol + provides the type information. 
		This definition basically takes two vectors as input, transforms them into values
		of type \hol{cfun}, computes their \hol{cfun}-addition, and transforms back
		the result into a vector.
		This operation can actually be easily seen to be equivalent to \hol{cvector\_add} in Definition \ref{def:cvector_add}.
		Therefore, not only we have a bijection between
		the types \hol{(N)finite\_image\to complex} and \hol{complex^N},
		but also an \emph{isomorphism} between the structures
		\hol{((N)finite\_image\to complex,+_{cfun})} and \hol{(complex^N,+_{complex^N})}.

We use these observations to develop a framework proving that there exists an isomorphism between the two type \hol{A \to complex} and \hol{complex^N} when \hol A has \hol N elements. This development results in a uniform library addressing both finite and infinite dimension vector spaces. The details of this development can be found in \cite{report_MCS}.

In order to show the effectiveness of our formalization, in the next section we formally describe monochromatic light waves and their behaviour at an interface between two mediums. Then we verify the laws of incidence and reflection. We intentionally developed our formalization very similar to what one can find in physics textbooks. 

  \section{Application: Monochromatic Light Waves}
  \label{sec:application}
In the electromagnetic theory, light is described by the same principles that govern all forms of electromagnetic radiations.
The reason we chose this application is that vector analysis provides an elegant mathematical language in which electromagnetic theory is conveniently expressed and best understood. 
In fact, in the literature, the two subjects, electromagnetic theory and complex vector analysis, are so entangled, that one is never explained without an introduction to the other. Note that, although physically meaningful quantities can only be represented by real numbers, in analytical approaches, it is usually more convenient to introduce the complex exponential rather than real sinusoidal functions. 

An electromagnetic radiation is composed of an electric and a magnetic field. 
The general definition of a field is ``a physical quantity associated with each point of space-time''.
Considering electromagnetic fields (``EMF''), the ``physical quantity'' consists of a 3-dimensional complex vector for the electric and the magnetic field.
Consequently, both those fields are defined as complex vector valued functions $\vec{E}(\vec{r},t)$ and $\vec{H}(\vec{r},t)$, respectively, where $\vec r$ is the position and $t$ is the time.
Points of space are represented by 3-dimensional real vectors, so we define the type \hol{point} as an abbreviation for the type \hol{{real}^3}. Instants of time are considered as real so the type \hol{time} represents type ${real}$ in our formalization.
Consequently, the type \hol{field} (either magnetic or electric) is defined as \hol{point \to time \to {complex}^3}.
	Then, since an EMF is composed of an electric and a magnetic field, we define the type \hol \emf\ to represent \hol{point \to time \to {complex}^3 \times {complex}^3}.
	The electric and magnetic fields are therefore complex vectors.
	Hence their formalization and properties make use of the complex vectors theory developed in Section \ref{sec:CVS}.

One very important aspect in the formalization of physics is to make sure that all the postulates enforced by physics are formalized. We call these sets of definitions as ``constraints''. For instance, we define a predicate which ensures that the electric and magnetic field of an electromagnetic field are always orthogonal, as follows: 

\begin{cons}[Valid Electromagnetic Field]\  \\
\hol {\vdash \Forall {emf}
 is\_valid\_emf\ emf \Leftrightarrow }\\
 \hol { \hspace{2cm} (\Forall{r\ t} corthogonal\ (e\_of\_emf\ emf\ r\ t)\ (h\_of\_emf\ emf\ r\ t))}
\end{cons}
 \noindent where \hol{e\_of\_emf} and \hol{h\_of\_emf} are two helpers returning the electric field and magnetic field of an \hol{emf}, respectively.

An electromagnetic plane wave can be expressed as
$\vec{U}(\vec{r},t) = \vec a(\vec{r}) e^{j\phi(\vec{r})} e^{j\omega t}$,
where $\vec{U}$ can be either the electric or magnetic field at point $\vec{r}$ and time $t$,
$\vec a(\vec{r})$ is a complex vector called the \emph{amplitude} of the field,
and $\phi (\vec{r})$ is a complex number called its \emph{phase}.
In monochromatic plane waves (i.e., waves with only one \emph{frequency} $\omega$), the phase $\phi(\vec{r})$ has the form $-\vec{k} \odot \vec{r}$,
where ``$\odot$'' denotes the inner product between real vectors.
We call $\vec k$ the \emph{wavevector} of the wave; intuitively, this vector represents the propagation direction of the wave.
This yields the following definition:
\begin{definition}[Monochromatic Plane Wave] 
 \label{def_planar}\ \vspace{.1cm}\ \\
\hol{\vdash\planewave (k:{real}^3)\ (\omega:{real})\ (E:{complex}^3)\ (H:{complex}^3) : \emf }\ \vspace{.1cm}\\
\hol{\indnt \eqdef \Lambda{(r:point)\ (t:time)} (e^{-ii(\vec k \odot \vec r - \omega t)} E, e^{-ii (\vec k\odot \vec r - \omega t)} H)}
\end{definition}

\noindent where, again we accompany this physical definition with its corresponding Constraint \ref{valid_wave}, \hol{is\_valid\_wave}, which ensures that a plane wave \hol{\vec U} is indeed an electromagnetic field and the wavevector is indeed representing the propagation direction of the wave. This former condition will be satisfied, if and only if, the wavevector \hol k, electric field, and magnetic field of the wave are all perpendicular.  

\begin{cons}[Valid Monochromatic Wave]\ \label{valid_wave} \vspace{.1cm} \\
\hol {\vdash \Forall {emf}
is\_valid\_wave\ wave \Leftrightarrow }\\
 \hol {\hspace{1cm} ( is\_valid\_emf\ wave\ \wedge} \\
  \hol {\hspace{1cm} (\Exists{k\ w\ e\ h} } \\
  \hol {\hspace{1.2cm}\&0<w\ \wedge\ \neg(k = vec\ 0)\ \wedge\ wave = plane\_wave\ k\ w\ e\ h\ \wedge } \\
   \hol {\hspace{1.2cm} corthogonal\ e\ (vector\_to\_cvector\ k)\ \wedge} \\
   \hol{\hspace{1.2cm} corthogonal\ h\ (vector\_to\_cvector\ k)) }
\end{cons}

 \noindent where, \hol{vector\_to\_cvector} is a function from \hol{real^N} to \hol{complex^N}, mapping a real vector to a complex vector with no imaginary part.
 
Now, focusing on electromagnetic optics, when a light wave passes through a medium, its behaviour is governed by different characteristics of the medium.
The \emph{refractive index}, which is a real number, is the most dominant among these characteristics,
therefore the type \hol{medium} is defined simply as the abbreviation of the type ${real}$.
The analysis of an optical device is essentially concerned with the passing of light from one medium to another, hence we also define
the plane \emph{interface} between the two mediums, as \hol{medium \times medium \times plane \times real^3},
i.e., two mediums, a plane (defined as a set of points of space), and an orthonormal vector to the plane,
indicating which medium is on which side of the plane.

\begin{figure}[h]
  \centering  
  \includegraphics[scale=1]{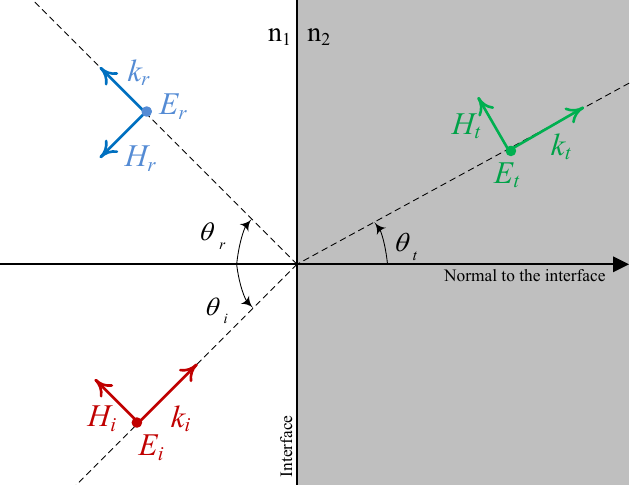}\ \\
  \caption{Plane Interface Between Two Mediums}\label{fig_interface}
  \vspace{-.2cm}
\end{figure}

In order to show how the effectiveness of our formalization, we prove some properties of waves at an interface
(e.g., the law of reflection, i.e., a wave is reflected in a symmetric way to the normal of the surface)
derived from the \emph{boundary conditions} on electromagnetic fields.
These conditions state that the projection of the electric and magnetic fields shall be equal on both sides of the interface plane.
This can be formally expressed by saying that the cross product between those fields and the normal to the surface shall be equal:

\begin{definition}[Boundary Conditions]\label{bound_cond}
\ \ \vspace{.2cm} \\ \hol{\vdashdef \boundaryconds emf_1\ emf_2\ n\ p\ t \Leftrightarrowdef}\ \\
\hol{\indnt n\times\eofemf emf_1\ p\ t = n\times \eofemf emf_2\ p\ t\  \wedge\ } \ \\
\hol{\indnt n\times \hofemf emf_1\ p\ t = n\times \hofemf emf_2\ p\ t}\\
\end{definition}
We then formalize a plane interface between two mediums, in the presence of a plane wave, shown in Fig. \ref{fig_interface}, with the following predicate:

 \begin{cons}[Plane Wave and a Plane Interface]\label{is_plane_wave}
	\ \ \vspace{.2cm}\\
	\hol{\vdashdef \isplanewaveatitf i\ emf_i\ emf_r\ emf_t \Leftrightarrowdef}\ \\
  \hol{\indnt \isvaliditf i\ \wedge\ \isplanewave emf_i\ \wedge}\ \\
  \hol{\indnt \isplanewave emf_r\ \wedge\ \isplanewave emf_t\ \wedge}\ \\
	\hol{\indnt let\ (n_1,n_2,p,n) = i\ in}\ \\
	\hol{\indnt let\ (k_i,k_r,k_t) = \mapt \kofwave (emf_i,emf_r,emf_t)\ in}\ \\
	\hol{\indnt let\ (e_i,e_r,e_t)\ =\mapt (norm\ \circ\ \eofwave)\ (emf_i,emf_r,emf_t)\ in}\ \\
	\hol{\indnt let\ (h_i,h_r,h_t)\ =\mapt (norm\ \circ\ \kofwave)\ (emf_i,emf_r,emf_t)\ in}\ \\
	\hol{\indnt 0\leq (k_i\odot \normalofplan p)\ \wedge\ (k_r\odot \normalofplan p)\leq0\ \wedge\ } \ \\
	\hol{\indnt 0 \leq (k_t\odot \normalofplan p)\ \wedge}\ \\
	\hol{\indnt (\Forall {pt}\ pt\in p \Rightarrow} \ \\
	\hol{\indnt \ \Forall t\boundaryconditions (emf_i+emf_r)\ emf_t\ n\ pt\ t)\ \wedge}\ \\
	\hol{\indnt \Exists{k_0}\ norm\ k_i = k_0 n_1\ \wedge\ norm\ k_r = k_0 n_1\ \wedge\ norm\ k_t = k_0 n_2\ \wedge}\ \\
 	\hol{\indnt \Exists{\eta_0} h_i = e_in_1/\eta_0\ \wedge\ h_r = e_rn_1/\eta_0\ \wedge\ h_t = e_tn_2/\eta_0\ \wedge\ e_i \neq 0\ \wedge\ e_r\neq 0}\ 
\end{cons}

\noindent where \hol{\mapt f\ (x,y,z)=(f\ x,f\ y,f\ z)},
\hol{norm} denotes the norm of a complex vector (defined by using the inner product),
\hol{e\_of\_w} (shorthand for ``electric field of wave'') is a helper function allowing to retrieve the electric amplitude of a wave,
and \hol{k\_of\_w} allows us to retrieve the wavevector of a wave.
The predicate of Constraint \ref{is_plane_wave} takes an interface \hol i and three EMFs \hol{e_i}, \hol{e_r}, and \hol{e_t},
intended to represent the incident wave, the reflected wave, and the transmitted wave, respectively.
The first four atoms of the predicate ensure that the arguments are well-formed,
i.e., \hol i is a valid interface and the three input fields exist and are plane waves
(we refer to the implementation for details about these predicates).
From this predicate, which totally describes the interface in Fig. \ref{fig_interface}, we can prove several foundational properties of plane waves.
For instance, the incident, reflected, and transmitted waves all lie in the same plane (the so-called ``plane of incidence''):
 \begin{theorem}[Law of Plane of Incidence] \ \vspace{.2cm}\\
	\label{thm_incidence}
  \hol{\vdash \Forall{i\ emf_i\ emf_r\ emf_t\ x\ y\ z}}\ \\
  \hol{\hspace{1cm} \isplanewaveatitf i\ emf_i\ emf_r\ emf_t\ \wedge\ } \ \\
  \hol{\hspace{1cm} \isincidentbasis (x,y,z)\ emf_i\ i\ \Rightarrow}\ \\
  \hol{\hspace{1cm} \kofwave emf_i\odot x = 0 \  \wedge \  \kofwave emf_r\odot x = 0 \ \wedge\ \ \kofwave emf_t\odot x = 0}
\end{theorem}
\noindent where \hol{\isincidentbasis\ (x,y,z)\ emf_i\ i} asserts that \hol{(x,y,z)} is a basis of the incident plane,
i.e., the plane characterized by the wavevector of \hol{emf_i} and the normal to \hol i
(note that if these two vectors are collinear then there is an infinity of planes of incidence).
Another non-trivial consequence is the fact that the reflected wave is symmetric to the incident wave with respect to the normal to the surface:
 \begin{theorem}[Law of reflection] \ \vspace{.2cm}\\ 
	\label{thm_reflection}
  \hol{\vdash \Forall{i\ emf_i\ emf_r\ emf_t}}\ \\
  \hol{\hspace{1cm} \isplanewaveatitf i\ emf_i\ emf_r\ emf_t\ \Rightarrow}\ \\
  \hol{\hspace{2.5cm} \aresymetric (-(\kofwave emf_i))\ (\kofwave emf_r)}\ \\
  \hol{\hspace{2.5cm}(\normalofplan(\planofitf i))}
\end{theorem}
\noindent where \hol{\aresymetric u\ v\ w} formalizes the fact that \hol u and \hol v  are symmetric with respect to \hol w
(this is easily expressed by saying that \hol{ v= 2*( u\odot  w) w - u}).
Referring to Fig. \ref{fig_interface}, Theorem \ref{thm_reflection} just means that $\theta_i=\theta_r$, which is the expression usually found in textbooks.

The proofs of these results make an extensive use of the formalization of complex vectors and their properties presented in Section \ref{sec:CVS}:
the arithmetic of complex vectors is used everywhere as well as the dot and cross product. Our proof script  for the application consists of approximately 1000 lines of code.
Without availability of the formalization of complex vectors, this reasoning would not have been possible, which indicates the usefulness of  our work.

  \section{Conclusion}
  \label{sec:conclusion}

	We presented formalization of  complex vectors in HOL Light. 
The concepts of linear spaces, norm, collinearity, orthogonality, vector angles, summability, and complex matrices which are all elements of more advanced concepts in complex vector analysis are formalized.
	An essential aspect of our formalization is to keep it \emph{engineering} and \emph{applied science}-oriented. 
Our libraries developed originally upon two different representations of complex vectors: vector of complex numbers (\hol{complex^N}) and bivectors (\hol{real^N \times real^N}), which were verified to be isomorphic.
	We favoured the notions and theorems that are useful in applications,
	and we provided tactics to automate the verification of most commonly encountered expressions.
We then illustrated the effectiveness of our formalization by formally describing electromagnetic fields and plane waves, 
	and we verified some classical laws of optics:
	the law of the plane of incidence and of reflection.

Our future works is extended in three different directions. The first is to enrich the libraries of complex vector analysis. We have not yet addressed differential and integral calculus of vector functions. Second, we plan to formally address practical engineering systems. 
  For instance, we already started developing electromagnetic applications,
  particularly in the verification of optical systems \cite{do-form-journal}.
  Finally, to have our formalization to be used by engineers, we need to develop more tactics and introduce automation to our libraries of complex vectors.

  \bibliographystyle{plain}
  \bibliography{cicm_ref}

\end{document}